\documentclass[useAMS,usenatbib,usegraphicx]{mn2e}

\usepackage{rotating}
\usepackage{lscape}

\title[Low mass X-ray binaries in the M31 globular cluster system] 
  {A systematic study of low mass X-ray binaries in the M31 globular cluster system}
\author[M. Peacock et al.]
{Mark B. Peacock$^{1}$\thanks{E-mail:mbp@soton.ac.uk (MBP)},
Thomas J. Maccarone$^{1}$, Arunav Kundu$^{2}$, Stephen E. Zepf$^{2}$\\
$^{1}$School of Physics and Astronomy, University of Southampton, Southampton, SO17 1BJ, UK\\
$^{2}$Department of Physics and Astronomy, Michigan State University, East Lansing, MI 48824, USA}

\begin{document}

\date{Released 2010 Xxxxx XX}

\pagerange{\pageref{firstpage}--\pageref{lastpage}} \pubyear{2010}

\maketitle

\label{firstpage}

\begin{abstract}
\label{sec:xray:abstract}

We investigate low mass X-ray binaries (LMXBs) in the M31 globular cluster (GC) system using data from the 2XMMi catalogue. These X-ray data are based on all publicly available \textit{XMM-Newton} observations of the galaxy. This new survey provides the most complete and homogeneous X-ray survey of M31's GCs to date, covering $>80\%$ of the confirmed old clusters in the galaxy. We associate 41 X-ray sources with confirmed old clusters in the M31 cluster catalogue of \citet{Peacock10}. Comparing these data with previous surveys of M31, it is found that three of these clusters are newly identified, including a bright transient source in the cluster B128. Four additional clusters, that are not detected in these 2XMMi data, have previously been associated with X-ray sources from \textit{Chandra} or \textit{ROSAT} observations. Including these clusters, we identify 45 clusters in M31 which are associated with X-ray emission. By considering the latest optical GC catalogues, we identify that three of the previously proposed X-ray clusters are likely to be background galaxies and two have stellar profiles. We consider the properties of LMXB-hosting clusters and confirm significant trends between the presence of an LMXB and the metallicity, luminosity and stellar collision rate of a cluster. We consider the relationship between the luminosity and stellar collision rate of a cluster and note that LMXB hosting clusters have higher than average stellar collision rates for their mass. Our findings strongly suggest that the stellar collision rate is the dominant parameter related to the presence of LMXBs. This is consistent with the formation of LMXBs in GCs through dynamical interactions. 

\end{abstract}


\section{Introduction}
\label{sec:xray:intro}

Globular clusters (GCs) are known to be a rich source of low mass X-ray binaries (LMXBs). In early studies of the Milky Way's X-ray population, it was realised that many more LMXBs were located in GCs than would be expected based on their masses alone. Of the $\sim$150 bright LMXBs currently known in the Milky Way, 14 reside in 12 of its GCs \citep{Liu01,Heinke09}. Since the Galactic GCs are thought to contain less than 0.1$\%$ of the stars in the Galaxy, this suggests that the formation of LMXBs is more than two orders of magnitude more efficient in GCs than in the field of the Galaxy. It has long been proposed that this extra efficiency is due to dynamical formation of these binaries in the dense cluster cores \citep{Katz75,Clark75}. The proposed mechanisms for dynamical formation of LMXB systems include: a donor star capturing the neutron star through tidal capture \citep[e.g.][]{Fabian75}; exchange interactions between a neutron star and a primordial binary system \citep{Hills76,Hut83}; or direct collisions of a neutron star with the envelope of a giant star \citep{Verbunt87}. These interactions are more likely to occur in the cores of GCs, than in field star populations, due to their high stellar densities. 

The properties of LMXB hosting clusters can be used to investigate the formation and evolution of these systems. For example, if dynamical formation is the primary method of forming LMXBs in GCs, then it is expected that there should be a direct relationship between the stellar collision rate in a cluster and the presence of an LMXB. Previous work has shown that the LMXBs in the Milky Way's GCs are consistent with dynamical formation scenarios \citep{Verbunt87}. However, studying such relationships in the Galactic GC system is limited by the low number of LMXBs in its GCs. In order to study more significant sample sizes in the Milky Way's GCs, it is necessary to combine different types of compact object binary systems \citep[e.g.][]{Pooley03}. 

In the era of \textit{Chandra} and \textit{XMM-Newton} it has become possible to study more significant samples of GC LMXB systems by looking at extragalactic sources. These observations of nearby galaxies confirm that their GCs also contain a large fraction of the galaxies' LMXBs. Unfortunately, investigating dynamical formation in extragalactic clusters is difficult due to the small angular sizes of typical cluster cores. However, relationships between the collision rate and presence of LMXBs have been suggested in M31 \citep{Peacock09}, Cen~A \citep{Jordan07} and possibly in M87 \citep{Jordan04,Waters07}. It is also found, both in the Milky Way and nearby galaxies, that LMXBs favour brighter (and hence more massive) GCs \citep[e.g.][]{Kundu02,Sarazin03,Kim06,Kundu07}. The likely reason for this is that higher mass clusters will generally have more stellar interactions and therefore form more LMXBs through dynamical interactions. However, it is also possible that LMXBs will favour high mass clusters because they may retain more of the neutron stars they produce. Neutron stars that are formed by core collapse may be formed with large kick velocities \citep[e.g.][]{Hobbs05}. In this case, the higher escape velocities of high mass clusters, may result in more neutron stars being retained by these clusters. However, it is also possible that neutron stars with lower kick velocities can be formed via electron capture \citep[e.g.][]{Pfahl02,ivanova08}. \citet{Smits06} demonstrated that the GC systems of six elliptical galaxies are consistent with the retention of neutron stars from a low kick velocity mode. 

Previous work on extragalactic LMXBs has also identified that metal rich clusters are more likely to host LMXBs than metal poor clusters \citep[e.g.][]{Bellazzini95,Kundu02,Kundu03}. Several explanations for this have been proposed. Metal rich stars are likely to be physically larger, which may result in more LMXBs forming through tidal interactions and direct collisions \citep{Bellazzini95}. It was shown by \citep{Maccarone04} that this effect alone is unlikely to explain the observed factor 3 enhancement of LMXBs in metal rich clusters. They propose that the irradiation induced winds in these binaries may explain the observed differences. These winds are likely to be stronger in metal poor systems due to decreased line cooling. \citet{Ivanova06} suggested that the metallicity relationship is a natural consequence of the properties of solar mass donor stars. In this mass range, they show that low metallicity stars lack an outer convective zone. This is likely to reduce the rate of tidal captures and also make it harder for a binary to tighten (and hence form an LMXB system). 

Here, we investigate LMXBs in M31's GC system. In section \ref{sec:xray:m31gc}, we consider the current optical and X-ray catalogues of M31's clusters. Section \ref{sec:xray:m31gc_xray} presents the X-ray properties of M31's clusters, based on \textit{XMM-Newton} observations of the galaxy, and compares these data with previous X-ray catalogues to investigate transient sources and potential contamination in these previous studies. Finally, in section \ref{sec:xray:lmxb_gc_properties}, we investigate the properties of clusters which are found to host LMXBs.

\section{M31 globular clusters} 
\label{sec:xray:m31gc}

The M31 GC system has been the focus of many studies. Despite this work, it is likely that some contamination and incompleteness currently exists in the M31 cluster catalogues. To identify clusters in M31 we use the recent catalogue of \citet[][hereafter P10]{Peacock10}. This catalogue includes all clusters and candidates identified in most major studies of the M31 GC system \citep[including those of: ][]{Battistini87,Barmby00,Galleti04,Kim07,Caldwell09}. It provides updated locations and classifications for all of these clusters. The locations of these clusters are found to be in good agreement with those of \citet{Caldwell09}, but are more accurate than those used previously to match with X-ray catalogues \citep[e.g.][]{Barmby00,Galleti04}. The work of P10 and \citet{Caldwell09} has also identified significant contamination in the previous catalogues from stellar sources, background galaxies and young clusters. 

M31 has been extensively surveyed by most recent X-ray observatories including: \textit{Einstein} \citep[e.g.][]{Trinchieri91}; \textit{ROSAT} \citep[e.g.][]{Supper97,Supper01}; \textit{XMM-Newton} \citep[e.g.][]{Shirey01,Trudolyubov04} and \textit{Chandra} \citep[e.g.][]{Kaaret02,Williams04}. Many of the resulting X-ray source catalogues have attempted to identify which sources are associated with GCs. \citet{Supper01} associated 33 sources in their \textit{ROSAT} survey with known GCs from the combined GC catalogues of \citet{Battistini87}, \citet{Battistini93} and \citet{Magnier93}. \textit{Chandra} observations were used by \citet{DiStefano02}, \citet{Kong02}, \citet{Kaaret02} and \citet{Williams04} to identify 28, 25, 25 and 26 GC X-ray sources, respectively. Currently the most complete M31 GC X-ray catalogue was produced by \citet{Trudolyubov04}. They combined \textit{XMM-Newton} observations along the disk of M31 with archived \textit{Chandra} observations, to investigate 43 X-ray sources which they associated with GCs \citep[from the GC catalogues of:][]{Battistini87,Magnier93,Barmby01,Galleti04}. Most recently, \citet{Fan05} collated the results of these previous studies to identify 54 unique GCs associated with X-ray sources. 

In this study, we consider the X-ray properties of M31's GCs using all publicly available \textit{XMM-Newton} observations of the galaxy. Since the study of \citet{Trudolyubov04}, the \textit{XMM-Newton} coverage of M31 has increased significantly and now covers the entire D$_{25}$ ellipse of the galaxy. These data provide more accurate source locations than the \textit{ROSAT} study of \citet{Supper01} and cover many more clusters than the previous X-ray studies of M31's GCs \citep{DiStefano02,Trudolyubov04}. In this study we restrict our analysis to only those clusters classed as confirmed clusters in the recent catalogue of M31 GCs of P10. The previous GC X-ray associations have generally considered all cluster candidates in the galaxy. While the presence of an X-ray source in a cluster does increase the probability of it being a genuine cluster, the inclusion of such sources is also likely to increase contamination.

\section{The X-ray population of M31's globular clusters} 
\label{sec:xray:m31gc_xray}

\subsection{\textit{XMM-Newton} observations of M31}

\begin{table*}
\centering
\begin{minipage}{140mm}
\caption{XMM-Newton observations of M31 \label{tab:2xmm_survey}}
\end{minipage}
\begin{minipage}{140mm}
\scriptsize
\begin{tabular}{@{}llrrrrrrc@{}}
\hline
\hline
Obs. date  & Target name & Rev no. & Obs. ID   & RA (deg) & DEC (deg) & PN exp (s) & M1 exp (s) & survey\\
\hline
2000-06-25 & M31Core     & 100     & 112570401 & 10.65    & 41.28     & 26.6       & 31.8       & n\\
2000-12-28 & M31Core & 193 & 112570601 & 10.71 & 41.24 & 9.8  & 12.2 & n\\
2001-01-11 & G1      & 200 & 65770101  & 8.23  & 39.56 & 5.0  & 7.4  & y\\
2001-06-29 & M31Core & 285 & 109270101 & 10.65 & 41.28 & 30.8 & 46.7 & n\\
2002-01-05 & M31North1 & 380 & 109270701 & 11.03 & 41.58 & 54.8 & 57.3 & y\\
2002-01-06 & M31Core & 381 & 112570101 & 10.71 & 41.25 & 61.1 & 63.6  & y\\
2002-01-12 & M31South1 & 384 & 112570201 & 10.39 & 40.91 & 56.1 & 58.1  & y\\
2002-01-24 & M31South2 & 390 & 112570301 & 10.06 & 40.58 & 38.1 & 51.2 & n\\
2002-01-26 & M31North2 & 391 & 109270301 & 11.37 & 41.92 & 27.3 & 32.5 & n\\
2002-06-29 & M31North3 & 468 & 109270401 & 11.63 & 42.29 & 52.7 & 57.1  & y\\
2003-02-06 & M31-Halo1 & 579 & 151581101 & 8.44 & 39.51 & 8.0 & 9.7 & n\\
2003-02-06 & M31-Halo4 & 579 & 151580401 & 11.57 & 41.34 & 11.3 & 13.1 & n\\
2003-07-01 & M31-Halo2 & 652 & 151581201 & 10.77 & 39.82 & 1.6 & 3.1 & n\\
2003-07-01 & M31-Halo3 & 652 & 151581301 & 11.47 & 40.73 & 1.9 & 4.4 & n\\
2004-01-02 & NGC205 & 745 & 204790401 & 10.12 & 41.67 & 10.7 & 12.9 & n\\
2004-07-16 & RXJ0042.6+4115 & 843 & 202230201 & 10.63 & 41.28 & 18.3 & 19.5 & n\\
2004-07-19 & RXJ0042.6+4115 & 844 & 202230501 & 10.63 & 41.28 & 9.1 & 13.1 & n\\
2004-07-19 & RXJ0042.6+4115 & 844 & 202230401 & 10.63 & 41.28 & 14.3 & 18.5 & n\\
2005-08-01 & 5C3.1 & 1034 & 300910201 & 10.44 & 40.37 & 8.7 & 3.1 & n\\
2006-06-28 & M31S3 & 1200 & 402560101 & 9.69 & 40.27 & 8.6 & 16.0  & y\\
2006-07-01 & M31SN1 & 1201 & 402560301 & 10.15 & 41.32 & 47.3 & 54.8  & y\\
2006-07-01 & M31SS1 & 1201 & 402560201 & 10.84 & 40.94 & 16.4 & 30.0  & y\\
2006-07-03 & M31nc2 & 1202 & 404060201 & 11.3 & 40.98 & 21.1 & 29.0 & n\\
2006-07-08 & M31SS2 & 1205 & 402560401 & 10.54 & 40.64 & 25.3 & 45.5  & y\\
2006-07-08 & Bo375 & 1205 & 403530201 & 11.41 & 41.68 & 0.0 & 5.3 & n\\
2006-07-10 & Bo375 & 1206 & 403530301 & 11.41 & 41.68 & 0.0 & 10.5 & n\\
2006-07-12 & Bo375 & 1207 & 403530401 & 11.41 & 41.68 & 0.0 & 9.9 & n\\
2006-07-14 & Bo375 & 1208 & 403530501 & 11.41 & 41.68 & 0.0 & 10.0 & n\\
2006-07-16 & Bo375 & 1209 & 403530601 & 11.41 & 41.68 & 0.0 & 12.7 & n\\
2006-07-20 & M31SN2 & 1211 & 402560501 & 9.89 & 40.99 & 52.9 & 58.6  & y\\
2006-07-23 & M31SN3 & 1212 & 402560701 & 9.73 & 40.64 & 25.1 & 56.1  & y\\
2006-07-28 & M31SS3 & 1215 & 402560601 & 10.16 & 40.36 & 31.9 & 37.8  & y\\
2006-12-25 & M31S2 & 1290 & 402560801 & 10.05 & 40.57 & 48.9 & 51.4  & y\\
2006-12-26 & M31NN1 & 1291 & 402560901 & 10.5 & 41.59 & 44.9 & 51.9  & y\\
2006-12-30 & M31NS1 & 1293 & 402561001 & 11.19 & 41.18 & 51.9 & 55.9  & y\\
2007-01-01 & M31NN2 & 1294 & 402561101 & 10.82 & 41.9 & 47.8 & 50.8  & y\\
2007-01-01 & M31NS2 & 1294 & 402561201 & 11.46 & 41.51 & 40.6 & 43.3  & y\\
2007-01-03 & M31NN3 & 1295 & 402561301 & 11.22 & 42.14 & 36.1 & 38.6  & y\\
2007-01-03 & M31NS3 & 1295 & 402561401 & 11.69 & 41.88 & 45.2 & 50.0  & y\\
2007-01-05 & M31N2 & 1296 & 402561501 & 11.36 & 41.92 & 43.7 & 46.2  & y\\
2007-07-25 & CXOM31J004059.2+41 & 1396 & 410582001 & 10.21 & 41.27 & 13.3 & 17.2 & n\\
\hline
\end{tabular}\\
\footnotesize
All \textit{XMM-Newton} observations of M31 which are included in the 2XMMi catalogue. The final column indicates whether the observation is considered part of our `homogeneous' survey of M31's GCs. The coordinates indicate the approximate centre of each observation. 
\end{minipage}
\end{table*}

Over the past decade, M31 has been the target of several \textit{XMM-Newton} observations. The first 10 observations of M31 were obtained as part of the science verification of the telescope. These observations included five observations along the disk of the galaxy, four observations of the core of the galaxy and a shorter observation of the halo GC G1. These data are described by \citet{Trudolyubov04}, who use it for their study of the galaxy's GCs. Since these initial observations, several other fields of the galaxy have been observed. These include: repeated observations of the bright X-ray sources RXJ20042.6+4115 (a bright Z-source in M31) and B375 (the brightest GC X-ray source in the galaxy); an observation of the dwarf galaxy NGC~205; an observation on the minor axis of M31; and four observations to cover the recently discovered extended halo GCs in the galaxy. In addition to these observations a survey was recently completed to cover the entire D$_{25}$ ellipse of the galaxy. All of these observations are summarised in table \ref{tab:2xmm_survey}. 

The initial observations along the disk of the galaxy, combined with these new observations covering the outer regions, provide a relatively homogeneous survey across the entire D$_{25}$ ellipse of the galaxy. These observations, identified by the flag \textit{survey}=y in table \ref{tab:2xmm_survey}, provide the basis for our primary survey. It should be noted that, some of the LMXBs in M31's GCs are expected to be transient in nature. For this reason we also search for GC LMXBs in the other observations available. This allows us to identify as many LMXB hosting clusters as possible. However, it does bias the detection of transient LMXBs to regions which have been more frequently observed (such as the central region of the galaxy). We therefore flag in our final table whether the observed GC X-ray sources would have been identified in our primary dataset (which essentially gives a single epoch survey across M31). It can be seen from table \ref{tab:2xmm_survey} that the exposure times for the different fields do vary slightly. The homogeneous detection limit across this survey is L$_{x}\sim$10$^{36}$erg$/$s corresponding to the relatively short observation of G1. In total, the \textit{XMM-Newton} observations considered here cover $\sim$82$\%$ of the confirmed old GCs from P10. 

\subsection{The 2XMMi catalogue}

To identify sources in these observations we use the `incremental second \textit{XMM-Newton} serendipitous source catalogue' \citep[2XMMi: for details see e.g.][]{Watson09}. This is a pipeline produced catalogue of sources which are identified in over 4000 publically available \textit{XMM-Newton} observations \footnote{The catalogue was downloaded on 18/11/2009 from http$:$//xmmssc-www.star.le.ac.uk/Catalogue/2XMMi/}. The catalogue includes the location, flux and reliability of sources detected in all of the M31 observations listed in table \ref{tab:2xmm_survey}. 

\subsection{M31 globular clusters in 2XMMi} 
\label{sec:m31gc_2xmmi_data}

The confirmed old, young and candidate clusters from P10 were matched to the 2XMMi catalogue using a matching radius of 2$\arcsec$+1$\arcsec$+$\sigma_{\rm{pos,2XMMi}}$. This matching radius was chosen to cover the errors in the optical cluster locations (conservatively estimated to be $<$2$\arcsec$, to account for systematic and centring errors for these extended sources) and systematic errors on the X-ray locations ($<$1$\arcsec$). This identified X-ray sources in 42 old clusters and 2 candidate clusters. No X-ray sources were identified in the young clusters listed in this catalogue.  

We investigated chance associations between the clusters and X-ray sources by shifting the cluster locations by $\pm5\arcsec$. The four resulting cluster locations were matched to the X-ray sources using the same matching criteria. Only two associations were found from this, with a relatively large offset of $>3\arcsec$. We therefore do not believe that there is significant contamination from chance associations in the sample. Table \ref{tab:m31gc_2xmmi} lists all old and candidate clusters associated with X-ray sources in the 2XMMi catalogue. 

It can be seen from table \ref{tab:m31gc_2xmmi} that most of the clusters are well matched to their proposed X-ray counterparts, with separations of $<$2$\arcsec$. However, three of the clusters have relatively large offsets of $>$3$\arcsec$. Comparison with published \textit{Chandra} observations (discussed in the next section) confirms that two of these sources, B094 and B146, are associated with the clusters. Both the \textit{XMM} and \textit{Chandra} locations of the source associated with B094 are slightly offset from the centre of the cluster. This may potentially align with a faint optical source on the edge of this cluster. Since LMXBs are expected to reside in the cores of the clusters (where the density is highest), it is possible that this source could be a chance alignment with the cluster. However, given its proximity to the cluster, we keep this source in our analysis, but note that its association is relatively uncertain. It is found that the cluster B035 does not appear to be associated with the matched X-ray source. Another faint counterpart is identified in K-band images that is in good agreement with the X-ray source location. We therefore believe that this may be a chance alignment with a non-cluster object and remove it from our analysis. 

For each GC X-ray source, table \ref{tab:m31gc_2xmmi} includes hardness ratios for the source and its total luminosity ($L_{x}, 0.2-12$keV). These data are taken from the 2XMMi `slim' catalogue. This is a reduced version of the main catalogue and contains only unique sources. Many of these sources are expected to vary during the observations and some are located in more than one observation. For these sources, this catalogue quotes the mean locations and fluxes. The hardness ratios (HR) of a source are defined as: 

\begin{equation}
 {\rm HR}_{i}=\frac{({\rm B}_{i+1}-{\rm B}_{i})}{({\rm B}_{i+1}+{\rm B}_{i})}
 \label{eq:hardness}
\end{equation}
Here, B$_{i}$ are the narrow energy bands: B$_{1}$=0.2-0.5~keV; B$_{2}$=0.5-1.0~keV; B$_{3}$=1.0-2.0~keV; B$_{4}$=2.0-4.5~keV; B$_{5}$=4.5-12~keV. Table \ref{tab:m31gc_2xmmi} also includes the summary flag (SUMFLG) from the 2XMMi catalogue. This gives an indication of the reliability of the detection. The relevant flags \citep[taken from][]{Watson09} are: \\
0 = good \\
1 = source parameters may be affected \\
2 = possibly spurious \\
3 = located in a region where spurious detections may occur \\
It can be seen that many of the detections have non-zero warning flags. This is likely due to the crowded nature of M31, with most of the SUMFLG=3 sources near the centre of the galaxy. \citet{Watson09} suggest that sources with flags 0-2 should be genuine, although sources with flags 1 or 2 have some of the automated spurious detection flags set. Class 3 sources are confirmed by `manual' flagging. However, they have all of the 2XMMi automated detection flags set to spurious, and may be spurious detections. Of the 43 old clusters associated with X-ray sources, 13 have SUMFLG=3. We note, from comparison with previous observations, that 10 of these 13 GCs have already been identified independently from \textit{Chandra} observations. These sources are therefore unlikely to be spurious. The \textit{XMM~Newton} thumbnail images for all of the 2XMMi GC detections\footnote{from http$\:$//www.ledas.ac.uk/data/2XMMi/} were also visually examined to identify any artifacts (for example due to chip gaps). This examination suggested that the source in the cluster AU010 was unlikely to be reliable and we remove it from our catalogue.

\subsection{Detections from previous catalogues} 

In addition to the \textit{XMM-Newton} observations used here, other X-ray observations of M31 have associated X-ray sources with its GCs. This previous work is summarised in section \ref{sec:xray:m31gc}. Given the transient nature expected from some LMXBs, it is likely that they may only be detected at certain epochs. For this reason, we consider which clusters in table \ref{tab:m31gc_2xmmi} were previously detected and identify any additional clusters with proposed X-ray emission. 

All confirmed clusters were matched to sources identified from: \textit{ROSAT} observations by \citet{Supper01}; \textit{Chandra} observations by \citet{DiStefano02}, \citet{Kong02}, \citet{Kaaret02}, \citet{Williams04} and \textit{XMM-Newton} and \textit{Chandra} observations by \citet{Trudolyubov04}. The M31 cluster catalogue used in this study has more accurate locations, and some additional clusters, compared with those used in these previous studies. We therefore consider all X-ray sources in these previous catalogues. Table \ref{tab:m31gc_xray_all_old} lists all confirmed clusters which are associated with X-ray sources from the 2XMMi catalogue or these previous studies. It can be seen that 38 of the clusters identified in the 2XMMi catalogue were identified previously by one of these studies, while three are newly identified. 

\begin{landscape}
\begin{table}
\caption{Old and candidate clusters associated with X-ray sources in the 2XMMi catalogue}
\label{tab:m31gc_2xmmi}
\centering
\scriptsize
\begin{tabular}{@{}lclrrcrrccccccc@{}}
\hline
\hline
GC NAME$^{1}$ & f$^{1}$ & 2XMMi Source ID & RA & DEC & POS$_{err}$ & L$_{x}^{2}$ & L$_{x,err}$ & $HR_{1}^{3}$ & $HR_{2}^{3}$ & $HR_{3}^{3}$ & $HR_{4}^{3}$ & SUMFLG$^{4}$ & $\Delta$POS$^{5}$ & survey$^{6}$  \\
        &           &       & deg   & deg    & arcsec & 10$^{35}$erg/s & 10$^{35}$erg/s &   &   &   &  & arcsec & arcsec & \\
\hline
G001-MII & 1 & 2XMM J003246.5+393440 & 8.194110 & 39.577931 & 1.22 & 17.8 & 4.6 & 0.36 & 0.28 & -0.39 & -0.83 & 0 & 0.61 & y\\
B005-G052 & 1 & 2XMM J004020.1+404358 & 10.083819 & 40.732910 & 0.71 & 2528.7 & 26.5 & 0.57 & 0.33 & -0.17 & -0.30 & 3 & 2.19 & y\\
B024-G082 & 1 & 2XMM J004111.7+414547 & 10.299454 & 41.763552 & 0.76 & 12.2 & 4.2 & 0.40 & -0.00 & -0.52 & -0.16 & 1 & 0.53 & y\\
B045-G108 & 1 & 2XMMi J004143.1+413419 & 10.429738 & 41.572200 & 0.35 & 2317.5 & 20.0 & 0.59 & 0.32 & -0.14 & -0.41 & 3 & 0.38 & y\\
B050-G113 & 1 & 2XMMi J004146.3+413218 & 10.442963 & 41.538495 & 0.63 & 31.4 & 3.7 & 0.71 & 0.35 & -0.10 & -0.23 & 0 & 0.33 & y\\
B055-G116 & 1 & 2XMM J004150.3+411211 & 10.459870 & 41.203495 & 0.44 & 26.0 & 2.8 & 0.83 & 0.75 & 0.03 & -0.14 & 1 & 0.36 & y\\
B058-G119 & 1 & 2XMM J004152.9+404708 & 10.470541 & 40.786082 & 0.79 & 15.7 & 4.0 & 0.43 & -0.07 & -0.27 & 0.29 & 1 & 0.82 & y\\
MITA140 & 1 & 2XMM J004209.4+411744 & 10.539309 & 41.295845 & 0.27 & 82.3 & 2.8 & 0.87 & 0.49 & -0.29 & -0.52 & 1 & 0.88 & y\\
B078-G140 & 1 & 2XMM J004212.0+411757 & 10.550226 & 41.299417 & 0.27 & 98.6 & 3.0 & 0.90 & 0.64 & -0.19 & -0.41 & 1 & 1.57 & y\\
B082-G144 & 1 & 2XMM J004215.7+410114 & 10.565752 & 41.020601 & 0.21 & 2828.8 & 34.9 & 0.78 & 0.63 & -0.02 & -0.16 & 1 & 0.67 & y\\
B086-G148 & 1 & 2XMM J004218.6+411401 & 10.577680 & 41.233678 & 0.22 & 672.0 & 6.6 & 0.48 & 0.35 & -0.18 & -0.39 & 1 & 0.78 & y\\
B094-G156 & 1 & 2XMM J004224.9+405720 & 10.603903 & 40.955641 & 0.94 & 21.4 & 6.1 & 0.37 & 0.36 & -0.23 & -0.54 & 1 & 2.95 & y\\
B096-G158 & 1 & 2XMM J004225.9+411914 & 10.608139 & 41.320638 & 0.25 & 106.2 & 2.6 & 0.80 & 0.45 & -0.28 & -0.51 & 0 & 1.36 & y\\
B098 & 1 & 2XMM J004227.3+405936 & 10.613964 & 40.993580 & 1.01 & 16.3 & 6.1 & 0.47 & 0.13 & -0.27 & 0.18 & 0 & 0.95 & y\\
B107-G169 & 1 & 2XMM J004231.1+411938 & 10.629791 & 41.327346 & 0.24 & 231.6 & 3.6 & 0.65 & 0.37 & -0.25 & -0.46 & 0 & 1.26 & y\\
B110-G172 & 1 & 2XMM J004233.0+410328 & 10.637729 & 41.057884 & 0.31 & 57.3 & 4.3 & 0.48 & 0.35 & -0.35 & -0.27 & 2 & 0.55 & y\\
B117-G176 & 1 & 2XMMi J004234.3+405710 & 10.642969 & 40.952834 & 1.48 & 14.2 & 10.3 & 0.10 & 0.84 & -0.19 & -0.25 & 0 & 1.10 & y\\
B116-G178 & 1 & 2XMMi J004234.5+413251 & 10.644032 & 41.547694 & 0.39 & 561.0 & 19.8 & 0.73 & 0.66 & -0.02 & -0.37 & 0 & 0.49 & y\\
B123-G182 & 1 & 2XMM J004240.6+411033 & 10.669252 & 41.175928 & 0.44 & 14.0 & 1.4 & 0.70 & 0.26 & -0.43 & -0.16 & 2 & 0.44 & y\\
B124-NB10 & 1 & 2XMM J004241.3+411524 & 10.672199 & 41.256867 & 0.42 & 49.1 & 3.4 & 0.59 & 0.06 & -0.21 & -0.59 & 3 & 1.40 & y\\
B128-G187 & 1 & 2XMM J004247.6+411113 & 10.698577 & 41.187145 & 0.55 & 570.2 & 12.6 & 0.46 & 0.23 & -0.23 & -0.36 & 1 & 1.66 & n\\
BH18 & 1 & 2XMM J004250.8+411032 & 10.711702 & 41.175787 & 1.31 & 6.0 & 2.1 & 0.62 & 0.81 & -0.03 & -0.31 & 0 & 1.21 & y\\
B135-G192 & 1 & 2XMM J004251.9+413108 & 10.716459 & 41.519072 & 0.25 & 4775.9 & 45.9 & 0.69 & 0.50 & -0.11 & -0.34 & 3 & 0.48 & y\\
B143-G198 & 1 & 2XMM J004259.5+411919 & 10.748114 & 41.322081 & 0.22 & 395.2 & 4.3 & 0.46 & 0.20 & -0.30 & -0.53 & 1 & 1.05 & y\\
B144 & 1 & 2XMM J004259.8+411606 & 10.749205 & 41.268360 & 0.22 & 513.8 & 5.2 & 0.49 & 0.23 & -0.19 & -0.36 & 3 & 0.63 & y\\
B091D-D058 & 1 & 2XMM J004301.4+413017 & 10.755850 & 41.504758 & 0.24 & 860.8 & 21.8 & 0.48 & 0.43 & 0.09 & -0.11 & 2 & 0.40 & y\\
B146 & 1 & 2XMM J004303.0+411525 & 10.762521 & 41.257129 & 0.22 & 421.3 & 4.4 & 0.48 & 0.18 & -0.32 & -0.48 & 3 & 3.17 & y\\
B147-G199 & 1 & 2XMM J004303.2+412121 & 10.763346 & 41.355999 & 0.24 & 161.7 & 3.7 & 0.43 & 0.10 & -0.39 & -0.41 & 1 & 1.10 & y\\
B148-G200 & 1 & 2XMM J004303.7+411805 & 10.765745 & 41.301462 & 0.23 & 303.2 & 4.2 & 0.44 & 0.21 & -0.31 & -0.38 & 3 & 0.95 & y\\
B150-G203 & 1 & 2XMM J004307.3+412019 & 10.780786 & 41.338813 & 0.36 & 28.5 & 1.9 & 0.49 & 0.23 & -0.28 & -0.40 & 1 & 1.34 & y\\
B153 & 1 & 2XMM J004310.5+411451 & 10.793951 & 41.247775 & 0.22 & 1176.8 & 8.0 & 0.47 & 0.21 & -0.21 & -0.36 & 3 & 0.94 & y\\
B158-G213 & 1 & 2XMM J004314.3+410721 & 10.809660 & 41.122499 & 0.20 & 1381.5 & 14.7 & 0.43 & 0.42 & 0.19 & -0.02 & 3 & 0.82 & y\\
B159 & 1 & 2XMM J004314.4+412513 & 10.810355 & 41.420344 & 0.68 & 13.5 & 2.7 & 0.78 & 0.70 & -0.36 & -0.17 & 0 & 1.73 & y\\
B161-G215 & 1 & 2XMM J004315.4+411125 & 10.814321 & 41.190373 & 0.43 & 16.2 & 2.1 & 0.47 & 0.04 & -0.38 & -0.34 & 0 & 0.48 & y\\
B182-G233 & 1 & 2XMM J004336.6+410813 & 10.902912 & 41.136951 & 0.71 & 57.2 & 8.6 & 0.50 & 0.46 & -0.18 & -0.07 & 0 & 0.95 & y\\
B185-G235 & 1 & 2XMM J004337.1+411443 & 10.905050 & 41.245554 & 0.19 & 867.1 & 9.5 & 0.50 & 0.27 & -0.19 & -0.25 & 1 & 0.90 & y\\
B193-G244 & 1 & 2XMM J004345.4+413656 & 10.939501 & 41.615799 & 0.45 & 52.8 & 3.6 & 0.33 & 0.14 & -0.16 & -0.52 & 0 & 0.82 & y\\
B204-G254 & 1 & 2XMM J004356.3+412203 & 10.985118 & 41.367648 & 0.41 & 62.5 & 6.6 & 0.37 & 0.25 & -0.37 & -0.25 & 0 & 0.64 & y\\
B225-G280 & 1 & 2XMMi J004429.5+412136 & 11.123257 & 41.360027 & 0.36 & 1175.4 & 22.6 & 0.45 & 0.24 & -0.26 & -0.31 & 1 & 0.44 & y\\
B375-G307 & 1 & 2XMMi J004545.4+413941 & 11.439532 & 41.661614 & 0.28 & 7013.5 & 40.2 & 0.56 & 0.35 & -0.18 & -0.46 & 3 & 0.94 & y\\
B386-G322 & 1 & 2XMM J004626.9+420153 & 11.612469 & 42.031350 & 0.19 & 1573.8 & 18.5 & 0.54 & 0.27 & -0.17 & -0.27 & 3 & 0.24 & y\\
\hline
[B035 & 1 & 2XMMi J004132.8+413830 & 10.386671 & 41.641724 & 0.58 & 14.2 & 2.0 & 0.47 & 0.12 & -0.26 & -0.79 & 0 & 3.36 & y]\\
\hline
SK091C & 2 & 2XMMi J004057.1+402155 & 10.238103 & 40.365344 & 1.54 & 5.7 & 2.2 & 0.77 & -0.64 & 0.83 & -0.28 & 0 & 1.86 & y\\
SK182C & 2 & 2XMMi J004527.2+413253 & 11.363430 & 41.548188 & 0.35 & 143.0 & 7.2 & 0.67 & 0.41 & -0.22 & -0.36 & 1 & 1.23 & y\\
\hline
\end{tabular}\\
\begin{minipage}{236mm}
\footnotesize
\textit{Top:} Old clusters associated with X-ray sources in the 2XMMi catalogue. \textit{Bottom:} candidate clusters associated with X-ray sources in the 2XMMi catalogue. \textit{Middle:} the cluster B035, which may \textit{not} be associated with this source. Columns 3-11 are taken from the slim version of the 2XMMi catalogue. \\
$^{1}$Name and classification from \citep{Peacock10}. \\
$^{2}$X-ray Luminosity (0.2-12keV), assuming all sources to be at 780kpc \citep{McConnachie05}. \\
$^{3}$X-ray hardness ratios, HR$_{i}$=(B$_{i+1}$-B$_{i}$)/(B$_{i+1}$+B$_{i}$). Where B$_{i}$ are the narrow energy bands: B$_{1}$=0.2-0.5keV; B$_{2}$=0.5-1.0keV; B$_{3}$=1.0-2.0keV; B$_{4}$=2.0-4.5keV; B$_{5}$=4.5-12keV\\ 
$^{4}$The quality flag assigned to the source by the 2XMMi pipeline (as described in section \ref{sec:m31gc_2xmmi_data}, see \citet{Watson09} for details). \\ 
$^{5}$Offset between the optical and X-ray source locations. \\ 
$^{6}$Indicates whether the source was detected in our `primary survey' observations. 
\end{minipage}
\end{table}
\end{landscape}

\begin{landscape}
\begin{table}
\caption{Old clusters with proposed X-ray emission}
\label{tab:m31gc_xray_all_old}
\centering
\scriptsize
\begin{tabular}{@{}lrrrrrcrrrrrrrrrrr@{}}
\hline
\hline
GC NAME & f & RA & DEC &\multicolumn{2}{c}{2XMMi}        & mixed         & \multicolumn{2}{c}{ROSAT} & \multicolumn{8}{c}{Chandra} \\
  &  &   &   & (this study) & $\Delta$POS & T04/F05$^{1}$ & S01$^{2}$ & $\Delta$POS & D02$^{3}$ & $\Delta$POS & Ko02$^{4}$ & $\Delta$POS & Ka02$^{5}$ & $\Delta$POS & W04$^{6}$ & $\Delta$POS \\
 &  &  &  & (0.2-12~keV) & arcsec & (mixed) & (0.1-2.0~keV) & arcsec & (0.3-7~keV) & arcsec & (0.3-7~keV) & arcsec & (0.1-10~keV) & arcsec & (0.3-7~keV) & arcsec  \\
\hline
G001-MII & 1 & 8.19389 & 39.57791 & 17.8 & 0.61 & 6 & - & - & - & - & - & - & - & - & - & - \\
B005-G052 & 1 & 10.08462 & 40.73287 & 2528.7 & 2.19 & 1990 & 1127.3 & 1.82 & - & - & - & - & - & - & 1469.0 & 0.28 \\
B024-G082 & 1 & 10.29938 & 41.76369 & 12.2 & 0.53 &  & - & - & - & - & - & - & - & - & - & - \\
B045-G108 & 1 & 10.42961 & 41.57224 & 2317.5 & 0.38 & 450 & 1173.0 & 2.20 & - & - & - & - & - & - & - & - \\
B050-G113 & 1 & 10.44285 & 41.53846 & 31.4 & 0.33 &  & - & - & - & - & - & - & - & - & - & - \\
B055-G116 & 1 & 10.45994 & 41.20341 & 26.0 & 0.36 & 2.4-2.9 & - & - & - & - & - & - & - & - & - & - \\
B058-G119 & 1 & 10.47083 & 40.78602 & 15.7 & 0.82 & 1-24 & - & - & 13.0 & 1.17 & - & - & - & - & - & - \\
MITA140 & 1 & 10.53956 & 41.29600 & 82.3 & 0.88 & 40-114 & - & - & 117.0 & 1.92 & 83.0 & 0.52 & 55.0 & 1.33 & 62.0 & 0.50 \\
B078-G140 & 1 & 10.55068 & 41.29969 & 98.6 & 1.57 & 24-300 & - & - & 26.0 & 2.63 & 101.0 & 0.68 & 25.0 & 0.54 & 108.0 & 0.27 \\
B082-G144 & 1 & 10.56597 & 41.02069 & 2828.8 & 0.67 & 1735-2330 & 777.6 & 3.04 & 1197.0 & 0.61 & - & - & 838.0 & 1.48 & 747.0 & 0.37 \\
B086-G148 & 1 & 10.57773 & 41.23389 & 672.0 & 0.78 & 498-763 & 409.3 & 7.87 & 469.0 & 2.66 & 456.0 & 0.59 & 247.0 & 0.12 & 430.0 & 0.42 \\
B094-G156 & 1 & 10.60434 & 40.95489 & 21.4 & 2.95 & 19 & 26.0 & 5.44 & 17.0 & 1.47 & - & - & - & - & - & - \\
B096-G158 & 1 & 10.60863 & 41.32072 & 106.2 & 1.36 & 28-174 & - & - & 74.0 & 1.00 & 93.0 & 0.91 & 45.0 & 0.42 & 114.0 & 0.85 \\
B098 & 1 & 10.61408 & 40.99333 & 16.3 & 0.95 & 1-12 & - & - & 8.0 & 0.24 & - & - & - & - & - & - \\
B107-G169 & 1 & 10.63022 & 41.32748 & 231.6 & 1.26 & 56-290 & 268.9 & 4.96 & 44.0 & 1.10 & 114.0 & 0.35 & 115.0 & 0.05 & 115.0 & 0.32 \\
B110-G172 & 1 & 10.63793 & 41.05787 & 57.3 & 0.55 & 41-52 & 109.0 & 2.04 & - & - & - & - & - & - & - & - \\
B117-G176 & 1 & 10.64321 & 40.95259 & 14.2 & 1.10 & 14 & - & - & - & - & - & - & - & - & - & - \\
B116-G178 & 1 & 10.64390 & 41.54760 & 561.0 & 0.49 & 234 & 137.6 & 2.24 & - & - & - & - & - & - & 234.0 & 3.67 \\
B123-G182 & 1 & 10.66941 & 41.17595 & 14.0 & 0.44 & 16-27 & - & - & 28.0 & 2.51 & 14.0 & 1.16 & 11.0 & 1.11 & 20.0 & 0.62 \\
B124-NB10 & 1 & 10.67261 & 41.25663 & 49.1 & 1.40 & 34-137 & - & - & 11.0 & 1.89 & 25.0 & 0.11 & 39.0 & 0.15 & 35.0 & 0.54 \\
B128-G187 & 1 & 10.69919 & 41.18718 & 570.2 & 1.66 &  & - & - & - & - & - & - & - & - & - & - \\
BH18 & 1 & 10.71132 & 41.17596 & 6.0 & 1.21 & 3 & - & - & 5.0 & 1.47 & 5.0 & 0.20 & - & - & - & - \\
B135-G192 & 1 & 10.71653 & 41.51895 & 4775.8 & 0.48 & 3093-4009 & 1519.2 & 1.51 & - & - & - & - & 1679.0 & 0.78 & 2125.0 & 0.52 \\
B143-G198 & 1 & 10.74850 & 41.32206 & 395.2 & 1.05 & 152-555 & 298.4 & 6.20 & 279.0 & 1.49 & 315.0 & 0.45 & 259.0 & 0.11 & 364.0 & 0.29 \\
B144 & 1 & 10.74942 & 41.26829 & 513.8 & 0.63 & 216-512 & - & - & 268.0 & 1.97 & 255.0 & 0.28 & 266.0 & 0.14 & 285.0 & 0.07 \\
B091D-D058 & 1 & 10.75598 & 41.50481 & 860.8 & 0.40 & 553-692 & 166.4 & 4.71 & - & - & - & - & 225.0 & 2.32 & - & - \\
B146 & 1 & 10.76212 & 41.25630 & 421.3 & 3.17 & 74-414 & 731.7 & 1.80 & 218.0 & 1.68 & 269.0 & 0.53 & 237.0 & 0.15 & 302.0 & 0.33 \\
B147-G199 & 1 & 10.76375 & 41.35604 & 161.7 & 1.10 & 52-194 & 165.4 & 5.64 & 109.0 & 1.35 & 119.0 & 0.69 & 78.0 & 0.23 & 153.0 & 0.52 \\
B148-G200 & 1 & 10.76607 & 41.30136 & 303.2 & 0.95 & 105-418 & 504.8 & 6.50 & 183.0 & 1.10 & 221.0 & 0.50 & 209.0 & 0.46 & 233.0 & 0.40 \\
B150-G203 & 1 & 10.78128 & 41.33883 & 28.5 & 1.34 & 16-72 & - & - & 27.0 & 0.79 & 37.0 & 0.57 & 47.0 & 1.30 & 37.0 & 0.14 \\
B153 & 1 & 10.79421 & 41.24760 & 1176.8 & 0.94 & 373-1248 & 701.8 & 4.48 & 843.0 & 1.37 & 832.0 & 0.65 & 506.0 & 0.22 & 943.0 & 0.34 \\
B158-G213 & 1 & 10.80996 & 41.12253 & 1381.5 & 0.82 & 600-1880 & 251.6 & 5.54 & 91.0 & 4.89 & 197.0 & 4.39 & 366.0 & 0.31 & 483.0 & 0.73 \\
B159 & 1 & 10.81099 & 41.42040 & 13.5 & 1.73 & 2 & 24.5 & 6.84 & - & - & 12.0 & 0.63 & - & - & - & - \\
B161-G215 & 1 & 10.81421 & 41.19027 & 16.2 & 0.48 & 16-22 & - & - & 6.0 & 0.98 & 14.0 & 1.00 & - & - & - & - \\
B182-G233 & 1 & 10.90280 & 41.13670 & 57.2 & 0.95 & 46 & - & - & - & - & - & - & - & - & - & - \\
B185-G235 & 1 & 10.90534 & 41.24543 & 867.1 & 0.90 & 454-1981 & 418.1 & 6.74 & 183.0 & 1.56 & 425.0 & 0.49 & 364.0 & 1.95 & 350.0 & 0.29 \\
B193-G244 & 1 & 10.93962 & 41.61601 & 52.8 & 0.82 & 44 & 153.8 & 4.89 & - & - & - & - & - & - & 42.0 & 1.50 \\
B204-G254 & 1 & 10.98510 & 41.36747 & 62.5 & 0.64 & 37 & 105.3 & 2.52 & - & - & - & - & - & - & - & - \\
B225-G280 & 1 & 11.12316 & 41.35993 & 1175.4 & 0.44 & 1130 & 650.3 & 1.80 & - & - & - & - & - & - & - & - \\
B375-G307 & 1 & 11.43983 & 41.66175 & 7013.5 & 0.94 & 5148-10372 & 2936.4 & 5.59 & 2994.0 & 1.81 & - & - & - & - & 4967.0 & 0.89 \\
B386-G322 & 1 & 11.61255 & 42.03132 & 1573.8 & 0.24 & 1496 & 822.2 & 3.89 & - & - & - & - & - & - & - & - \\
B293-G011 & 1 & 9.08691 & 40.89363 & - & - & 5.1 & 13.3 & 9.54 & - & - & - & - & - & - & - & - \\
B163-G217 & 1 & 10.82346 & 41.46252 & - & - & 1-1010 & 377.9 & 4.96 & - & - & - & - & - & - & - & - \\
B164-V253 & 1 & 10.82553 & 41.20813 & - & - & 13 & 33.4 & 10.54 & - & - & - & - & - & - & - & - \\
B213-G264 & 1 & 11.01463 & 41.51075 & - & - & 201 & - & - & - & - & - & - & - & - & - & - \\
\hline
\end{tabular}\\
\begin{minipage}{236mm}
\footnotesize
All old clusters from P10 with proposed X-ray associations from this, or previous, surveys. The object positions are the optical locations from P10. Where available, we list the associated X-ray luminosity ($\times$10$^{35}$ergs/s) and offset from the cluster location for sources in: the 2XMMi catalogue (this study); $^{1}$ \citep{Trudolyubov04}/\citep{Fan05}; $^{2}$ \citep{Supper01}; $^{3}$ \citep{DiStefano02}; $^{4}$ \citep{Kong02}; $^{5}$ \citep{Kaaret02} and $^{6}$ \citep{Williams04}. \\  
\end{minipage}
\end{table}
\end{landscape}

\begin{landscape}
\begin{table}
\begin{minipage}{242mm}
\vspace{45mm}
\caption{Candidate clusters and potentially misclassified clusters with proposed X-ray emission}
\label{tab:m31gc_xray_all_non}
\centering
\scriptsize
\begin{tabular}{@{}lrrrrrcrrrrrrrrrrr@{}}
\hline
\hline
GC NAME & f & RA & DEC &\multicolumn{2}{c}{2XMMi}        & mixed         & \multicolumn{2}{c}{ROSAT} & \multicolumn{8}{c}{Chandra} \\
  &  &   &   & (this study) & $\Delta$POS & T04/F05$^{1}$ & S01$^{2}$ & $\Delta$POS & D02$^{3}$ & $\Delta$POS & Ko02$^{4}$ & $\Delta$POS & Ka02$^{5}$ & $\Delta$POS & W04$^{6}$ & $\Delta$POS \\
 &  &  &  & (0.2-12keV) & arcsec & (mixed) & (0.1-2.0keV) & arcsec & (0.3-7keV) & arcsec & (0.3-7keV) & arcsec & (0.1-10keV) & arcsec & (0.3-7keV) & arcsec  \\
\hline
SK091C & 2 & 10.23810 & 40.36534 & 5.7 & 1.86 & - & - & - & - & - & - & - & - & - & - & - \\
SK182C & 2 & 11.36383 & 41.54835 & 143.0 & 1.23 & 15 & 39.5 & 12.91 & - & - & - & - & - & - & - & - \\
NB63 & 2 & 10.63016 & 41.33662 & - & - & 5 & - & - & - & - & 5.0 & 3.55 & - & - & - & - \\
BH16 & 2 & 10.69204 & 41.29333 & - & - & 9 & - & - & 31.0 & 1.41 & 9.0 & 0.43 & - & - & - & - \\
B138 & 2 & 10.73165 & 41.30981 & - & - & 8-84 & - & - & - & - & 20.0 & 0.37 & 23.0 & 0.12 & 35.0 & 0.19 \\
B007 & 4 & 10.10750 & 41.48667 & 72.5 & 16.05 & 71.5 & 187.4 & 20.4 & - & - & - & - & - & - & - & - \\
B042D & 4 & 10.52536 & 41.04669 & 62.4 & 0.40 & 31-73 & 42.4 & 8.29 & 85.0 & 0.18 & - & - & - & - & 42.0 & 0.69 \\
B044D-V228 & 4 & 10.52963 & 41.00458 & 158.3 & 0.78 & 40-100 & 45.2 & 2.52 & 71.0 & 0.50 & - & - & - & - & 38.0 & 1.51 \\
SK059A & 5 & 10.79109 & 41.31692 & 14.4 & 0.47 & 212 & 615.3 & 4.17 & - & - & 74.0 & 0.33 & 212.0 & 0.47 & 70.0 & 0.96 \\
B063D & 6 & 10.64586 & 40.81090 & 155.0 & 0.46 & 154 & 85.4 & 5.80 & - & - & - & - & - & - & 154.0 & 1.25 \\
SK119B & 6 & 10.77379 & 41.26622 & - & - & 43 & - & - & - & - & - & - & - & - & - & - \\
MIT311$^{*}$ & 0 & 10.92917 & 41.48111 & 58.8 & 2.81 & 52 & 55.5 & 2.1 & - & - & - & - & - & - & - & - \\
MIT380$^{*}$ & 0 & 11.10667 & 41.60778 & 21.1 & 6.42 & 27 & 71.2 & 5.8 & - & - & - & - & - & - & - & - \\
MIT16$^{*}$ & 0 & 10.13500 & 40.55778 & 14.9 & 14.83 & 9.0 & 23.6 & 3.8 & - & - & - & - & - & - & - & - \\
MIT317$^{*}$ & 0 & 10.9525 & 41.46278 & 14.3 & 15.86 &  & 43.2 & 39.9 & - & - & - & - & - & - & - & - \\
MIT165/166$^{*}$ & 0 & 10.58167 & 41.36472 & - & - & 5 & - & - & 8.0 & - & - & - & - & - & - & - \\
\hline
\end{tabular}\\
\end{minipage}
\begin{minipage}{236mm}
\footnotesize
Candidate clusters and (previously misclassified) non-clusters with proposed X-ray associations. The object positions are the optical locations from P10. The object classification flag (f) is taken from P10. Flags are: 0, no classification from P10; 1, old cluster; 2, unconfirmed candidate; 4, background galaxies; 5, HII region; 6, stellar object.  Where available, we list the associated X-ray luminosity ($\times$10$^{35}$ergs/s) and offset from the cluster location for sources in: the 2XMMi catalogue (this study); $^{1}$ \citep{Trudolyubov04}/\citep{Fan05}; $^{2}$ \citep{Supper01}; $^{3}$ \citep{DiStefano02}; $^{4}$ \citep{Kong02}; $^{5}$ \citep{Kaaret02} and $^{6}$ \citep{Williams04}. \\
$^{*}$Clusters from the study of \citet{Magnier93} are not listed in P10 (f=0), their names and locations are taken from \citet{Supper01} or \citet{Trudolyubov04}. 
\end{minipage}
\end{table}
\end{landscape}

We identify four GC X-ray sources in the previous work which are not identified in the 2XMMi catalogue. The cluster B163 was previously detected by \textit{ROSAT} and  by one of the \textit{Chandra} observations of it. This cluster is discussed by \citet{Trudolyubov04}, where they demonstrate its transient nature. The cluster B213 was detected in a deep \textit{ROSAT} HRI observation of the centre of M31 by \citet{Primini93}. The two other clusters, B164 and B293, were identified in the \textit{ROSAT} PSPC survey of \citet{Supper01}. These sources have relatively large offsets from the cluster locations of 10.5 and 9.5$\arcsec$, respectively. One of these sources is outside the region covered by the \textit{XMM} observations. The other three have no 2XMMi counterparts within the X-ray positional error. Since these sources have no 2XMMi associations, they are likely to be transient in nature. These clusters are included in our catalogue of X-ray clusters. However, it should be noted that the larger size of the \textit{ROSAT} positional error increases the probability of a chance association. 

\subsection{Candidate clusters and other sources} 
\label{sec:m31gc_2xmmi_previous}

As discussed in section \ref{sec:xray:m31gc}, there has been significant work in classifying and studying the properties of M31's clusters since the previous X-ray cluster catalogue of \citet{Trudolyubov04}. In this section, we consider the current classifications of previously proposed clusters. Table \ref{tab:m31gc_xray_all_non} lists all X-ray sources associated with unconfirmed candidate clusters and previously proposed X-ray clusters which have subsequently been reclassified. 

Of the 33 \textit{ROSAT} sources associated with clusters by \citet{Supper01}, 26 are associated with confirmed old clusters from P10. Two of the sources are now known to be background galaxies (B007 and B042D) and one is an unconfirmed candidate cluster. The other four clusters are from the catalogue of \citet{Magnier93} and do not have new classifications from P10. We believe that these clusters are not confirmed through spectroscopy or high resolution imaging. We therefore consider them as candidate (rather than confirmed) clusters. 

The study of \citet{Trudolyubov04} identified 43 X-ray clusters in M31, 37 of which are confirmed as old clusters by P10. Two of the clusters from P10 (B138 and SK100C) and two from \citet{Magnier93} (MIT165/166 and MIT133) are currently unconfirmed candidate clusters. The other two proposed GC X-ray sources in this catalogue are now thought to be background galaxies (B042D and B044D). We also note that these two galaxies, and an additional unconfirmed cluster (BH16), are included in the catalogue of M31 GC X-ray sources of \citet{DiStefano02}. In addition to the GC X-ray sources in these catalogues, the collated work of \citet{Fan05} identifies three other GC X-ray sources which are not listed as clusters in the catalogue of P10. One of these, B063D, has been reclassified as stellar. This source is detected in these 2XMMi data and is associated with the object. This object may potentially be an unresolved background galaxy or possibly a foreground flare star. Another source (SK059A) has been reclassified as an HII region by \citet{Caldwell09}. This source represents an interesting object for follow up investigation to determine its true nature. Finally, the source ``WSB85/S3-14'' in \citet{Fan05} is detected in these XMM data, but has no counterparts in P10. Examination of SDSS images of this cluster location shows no optical counterpart within the X-ray error circle. These images are deep enough to detect the entire GC luminosity function at the distance of M31. We therefore believe that this source is unlikely to be associated with a cluster and exclude it from table \ref{tab:m31gc_xray_all_non}.

\subsection{LMXB population} 
\label{sec:xray:lmxbs}

We associate X-ray sources with $\sim$11$\%$ of the old clusters in M31. In addition to these clusters, it is very likely that some clusters which host transient LMXBs were not detected in these observations. When considering M31's total cluster system, it should also be noted that there is likely to be significant incompleteness in the current optical cluster catalogues; especially at the faint end of the GC luminosity function. Given this incompleteness, the total fraction of clusters hosting LMXBs in M31 appears to be consistent with that of the Milky Way ($\sim$8$\%$). 

\begin{figure}
 \centering
 \includegraphics[height=84mm,angle=270]{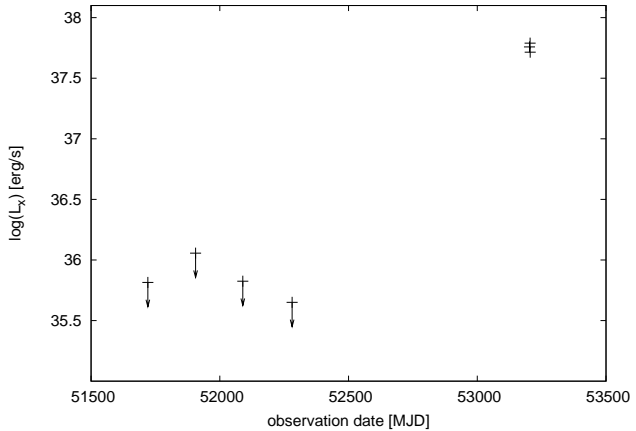}
 \caption{The transient X-ray source in B128.}
 \label{fig:B128}
\end{figure}

Only one of the cluster X-ray sources detected in these XMM observations would not have been detected in our `primary survey' observations (which essentially represents a single epoch of observations across the galaxy). This cluster (B128) was observed, but not detected, in three previous \textit{XMM-Newton} observations of the `core' of the galaxy. Figure \ref{fig:B128} shows the observations of this cluster. The upper limits in this plot are taken from the FLIX\footnote{http$:$//www.ledas.ac.uk/flix/flix.html} web tool and demonstrate that the source has increased in luminosity by over two orders of magnitude. As discussed above, four additional X-ray sources, that were previously identified in \textit{ROSAT} or \textit{Chandra} observations, were not detected in these XMM data and may also be transient sources. This demonstrates the benefit of re-observing clusters to identify LMXBs in outburst. In particular, elusive GC black hole LMXBs are likely to be transient in nature \citep[e.g.][]{King96}.  

\begin{figure}
 \centering
 \includegraphics[height=84mm,angle=270]{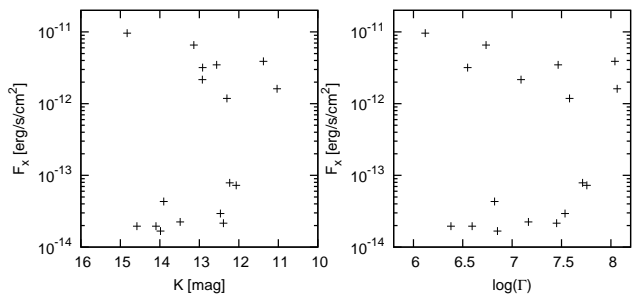}
 \caption{X-ray flux as a function of the clusters luminosity (left) and stellar collision rate (right).}
 \label{fig:Fx}
\end{figure}

\begin{figure*}
 \centering
 \includegraphics[height=176mm,angle=270]{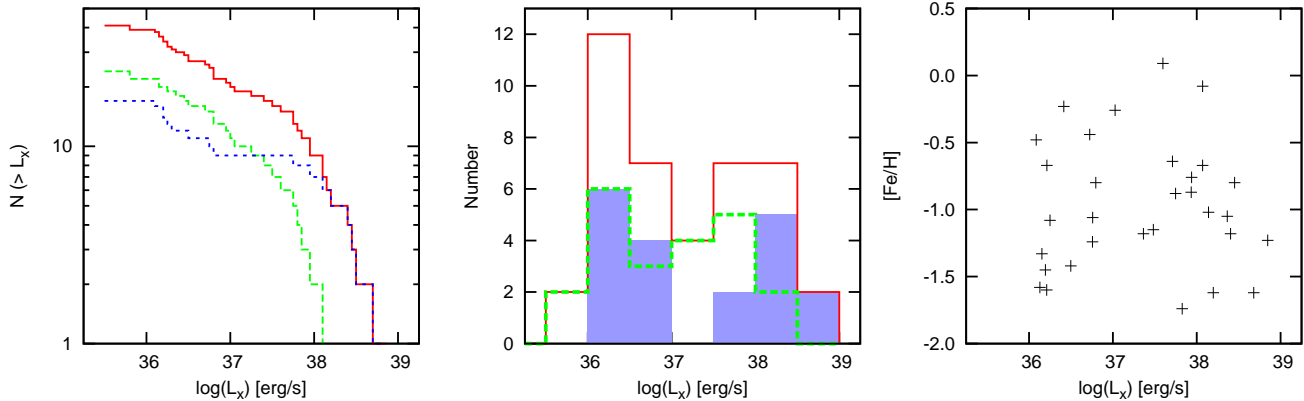}
 \caption{\textit{(left)} X-ray luminousity function of all clusters detected in this 2XMM survey (solid, red line), clusters with the $R_{{\rm gcp}}<3~{\rm kpc}$ (dashed, green line) and clusters with $R_{{\rm gcp}}>3~{\rm kpc}$ (dotted, blue line). In the outer clusters, the XLF appears to flatten at intermediate X-ray luminousities. \textit{(middle)} Histogram of X-ray luminousity of these clusters. This demonstrates the defeit of sources in the range $10^{37}<{\rm L}_{x}<10^{38}$ in the outer clusters (solid blue). \textit{(right)} X-ray luminousity as a function of metallicity (for the 31 LMXB hosting clusters with spectroscopic metallicities available). }
 \label{fig:hist_Lx}
\end{figure*}

To consider the spatial distribution of LMXB hosting clusters, we consider only those clusters covered by the observations selected as our primary survey. Of the 341 confirmed old clusters covered by these observations, 40 are associated with X-ray sources from our primary survey. It is found that the fraction of LMXB hosting clusters increases slightly towards the centre of the galaxy. However, it is also found that both the stellar collision rate and the metallicity of the clusters increases towards the centre of the galaxy. As shown in the next section, both of these parameters are expected to increase the formation of LMXBs. Selection effects in the most central region of the galaxy also bias the optical cluster catalogues against the detection of faint clusters (which are less likely to host LMXBs). As a result, this incompleteness may also produce such a relationship. For these reasons we believe these data do not provide evidence of a \textit{direct} relationship between galactocentric radius and the production of LMXBs. 

It can be seen from table \ref{tab:m31gc_2xmmi} that, compared with the Milky Way, a relatively large number of these LMXBs are very luminous, with $\sim15\%$ having L$_{x}>10^{38}$erg/s. Given that there are only 12 LMXB hosting GCs in the Milky Way, it is likely that the relative lack of systems this bright in the Milky Way may be related to the smaller sample size, rather than a difference between the two populations \citep[as proposed previously by e.g.][]{Supper97,Verbunt06}. It should also be noted that the LMXB in the Galactic GC M15 (AC~211) is an eclipsing (or near-eclipsing) source and does have an estimated intrinsic luminousity of L$_{x}>10^{38}$erg/s \citep{Naylor88}. We also note that LMXBs this bright are observed in the field of the Milky Way and in other extragalactic GCs \citep[e.g.][]{Liu01,Kundu02}. 

In principle, multiple unresolved LMXBs in M31's GCs could explain the high X-ray luminosities of some of these clusters. It is likely that some of these clusters will host more than one LMXB. For example, two of the 12 LMXB hosting clusters in the Milky Way are currently known to host two LMXBs in outburst \citep{White01,Heinke09}. However, variability observed in some of M31's LMXBs, suggests that one object is dominating the integrated luminosity. It can also be seen, from figure \ref{fig:Fx}, that we find no evidence of the X-ray luminosity increasing as a function of the stellar collision rate or the mass of a cluster. Such a relationship might be expected if the number of LMXBs in a cluster is $\gg1$. This is because the formation of LMXBs is known to be more efficient in these clusters (as discussed below). This suggests that, for these data, the variability in the luminosity of individual LMXBs appears to dominate over the effects of multiple LMXBs. 

Figure \ref{fig:Fx} suggests a potential bimodality in the X-ray luminousity function (XLF) of M31's GCs, with none of the clusters having X-ray fluxes (${\rm F}_{x}$) in the range $10^{-13}<{\rm F}_{x}<10^{-12}$ (compared with seven in the range $10^{-12}<{\rm F}_{x}<10^{-11}$ and nine in the range $10^{-14}<{\rm F}_{x}<10^{-13}$). If we assume a linear XLF for these clusters, then there is a low probability of 0.1$\%$ that none of the 17 clusters lie in this range. However, figure \ref{fig:Fx} includes only those clusters with stellar collision rates from \citet{Peacock09}. As such, it represents a subset of all the clusters in M31 which excludes clusters projected against the central (and outer halo) regions of M31. In the left and middle panels of figure \ref{fig:hist_Lx}, we plot the XLF for all clusters detected in this 2XMM survey (solid red line). We also split this sample into those clusters with projected galactocentric radius, $R_{{\rm gcp}}<3 {\rm kpc}$ (green, dashed line) and those with $R_{{\rm gcp}}>3 {\rm kpc}$ (blue dotted line/solid blue). It can be seen that this deficeit of clusters is identified in clusters projected against the outer, but not the inner, regions of the galaxy. To test whether there is a significant difference in the XLF of the inner and outer clusters, we run a Kolmogorov-Smirnov (K-S) test between the two populations. The probability that the inner and outer samples are drawn from different populations is 90$\%$. We note that the cut chosen between the inner and outer clusters is relatively arbetary, however similar probabilities are obtained using $R_{{\rm gcp}}$=2,4~kpc. These data therefore suggest a potential difference in the XLF at low and high $R_{{\rm gcp}}$, but with a low significance. 

If genuine, the relationship between $R_{{\rm gcp}}$ and the XLF may be suggestive of a metallicity effect on the formation of LMXBs in a cluster. This is because the metal rich clusters in M31 are known to be more centrally concentrated than the metal poor clusters \citep[see e.g. figure 11 of][]{Perrett02}. \citet{Ivanova06} have previously proposed that the formation of LMXBs with main sequence donor stars (through tidal captures in close encounters) is unlikely to occur in metal poor clusters. If this is the case, then it is likely that the LMXBs observed in metal poor clusters have primarily evolved donor stars. Such sources are likely to be Z-sources with high L$_{x}\sim 10^{38}$erg/s (if the donors are giants/subgiants) or ultracompact sources which will have a relatively steep luminousity function \citep[if the donors are white dwarfs;][]{Bildsten04}. In contrast, metal rich clusters would be expected to host a mixture of these LMXBs and LMXBs with main sequence donor stars. Hence LMXBs in the metal rich clusters may have a more uniform distribution of L$_{x}$. 

In the right panel of figure \ref{fig:hist_Lx} we investigate the effect of a clusters metallicity on the X-ray luminousity. No significant difference is observed between the X-ray luminousity of the metal rich (with [Fe/H]$>$-0.7) and metal poor clusters. However, spectroscopic metallicities are not available for all of the clusters hosting LMXBs. All of the clusters with $R_{{\rm gcp}}>3$ have metallicity estimates, but only 14 of the 24 clusters in the inner region of M31 have metallicities. Given their small galactocentric radius, many of these clusters with unknown metallicity may be metal rich. It is therefore possible that the observed galactocentric radius effect on the XLF may be suggestive of a metallicity effect, although these data show no direct evidence of this. More conclusive evidence of a metallicity effect, on the X-ray luminousity of LMXBs, may become available from very deep X-ray observations of galaxies with a larger number of metal rich clusters.

\section{LMXB-GC relationships}
\label{sec:xray:lmxb_gc_properties}

\begin{figure}
 \includegraphics[height=84mm,angle=270]{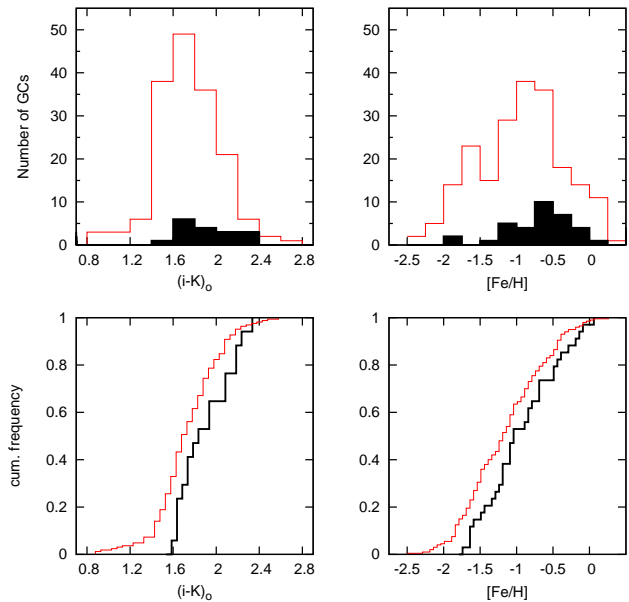}
 \caption{\textit{Top:} The dereddened colour and metallicity for all clusters (open, red) and LMXB hosting clusters (solid, black). \textit{Bottom:} the scaled cumulative frequency of the two populations. The cluster colours are taken from the catalogue of P10, while their spectroscopic metallicities taken from the collated table of \citet{Fan08}. It can be seen that the LMXBs tend to favour redder, higher metallicity, clusters.}
 \label{fig:lmxb_Z}
\end{figure}

\begin{figure*}
 \includegraphics[height=114mm,angle=270]{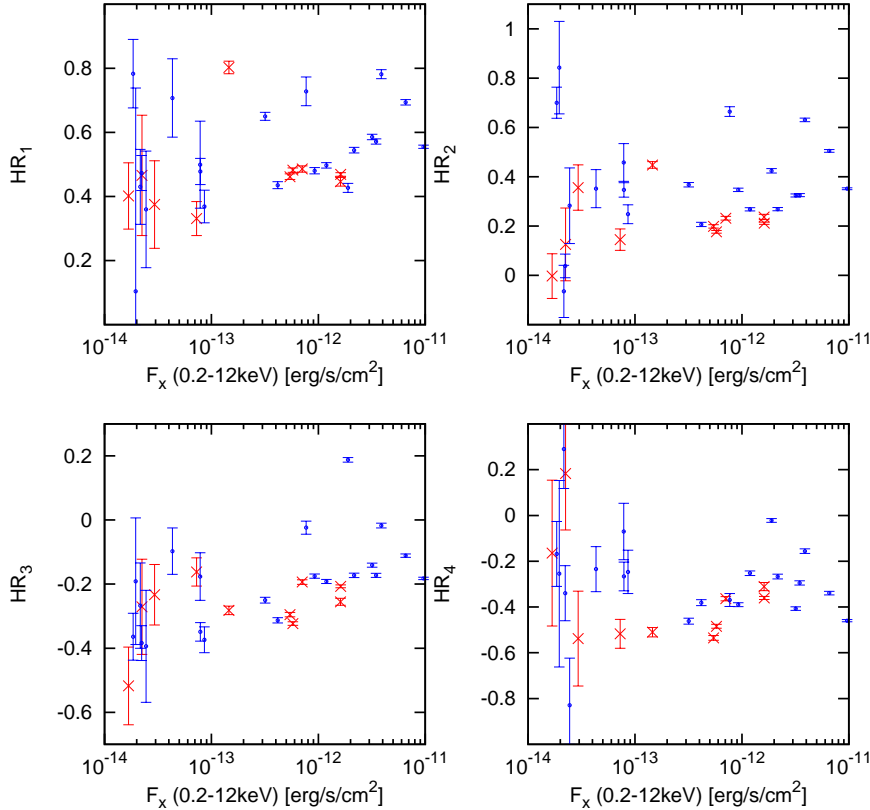}
 \caption{The hardness of a source as a function of its flux. The clusters are split into metal rich clusters (with $[Fe/H]<$-0.7, red crosses) and metal poor clusters (blue points). It can be seen that the metal rich clusters do appear to be softer, particularly in the HR$_{2}$.}
 \label{fig:lmxb_HR_Fx}
\end{figure*}

The properties of LMXB hosting clusters provide a useful insight into the effect of these properties on the formation and evolution of LMXBs. Figure \ref{fig:lmxb_Z} shows the spectroscopic metallicities for all clusters with data available from the collation of \citet{Fan08} \citep[from the studies of:][]{Huchra91,Barmby00,Perrett02}. These data cover $\sim50\%$ of the confirmed old clusters. The currently available spectroscopy for M31's clusters is biased towards the more massive (brighter) clusters. However, LMXBs are known to reside primarily in more massive clusters (as discussed below), so this work provides metallicities for 72$\%$ of the LMXB hosting clusters. We also consider the (\textit{i}-K) colour of these clusters from \citet{Peacock10}. These colours are dereddened using the values of \citet{Fan08} and provide an alternative estimation of the metallicity of the clusters. It can be seen from figure \ref{fig:lmxb_Z} that LMXBs favour redder, metal rich, clusters. A K-S test between the colour of all clusters and the LMXB hosting clusters suggests that there is a 98$\%$ likelihood that these clusters are drawn from different populations. For the spectroscopic metallicities this probability is 91$\%$. The statistical significance of this relationship is relatively weak, but is similar to that found in the previous studies of \citet{Bellazzini95} and \citet{Trudolyubov04}. A more significant metallicity effect has been previously observed in many other galaxies \citep[e.g.][]{Kundu02,Kim06}. There are several reasons why this relationship may be less obvious in M31. Firstly, there are relatively few metal rich clusters in M31 compared with many early type galaxies. It should also be noted that the spectroscopic metallicities of M31's clusters have relatively large errors (with a mean error of $\sim$0.25). The effect of this would be to weaken any genuine relationships between the clusters. In principle the colours of the clusters may give a more accurate measure of metallicity. However, the colours of M31's clusters are complicated due to variable extinction across the galaxy. 

\begin{figure}
 \includegraphics[height=84mm,angle=270]{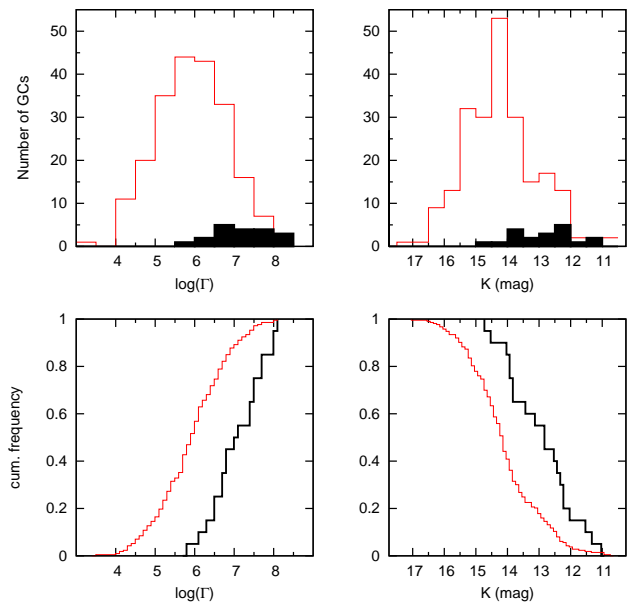}
 \caption{\textit{Top:} The stellar collision rate and luminosity for all clusters (open, red) and LMXB hosting clusters (solid, black). \textit{Bottom:} the scaled cumulative frequency of the two populations. The GC data are taken from the catalogue of P10. It can be seen that the LMXBs tend to favour brighter clusters with higher stellar collision rates.}
 \label{fig:lmxb_gamma_K}
\end{figure}

It has also been proposed that decreasing metallicity may lead to hardening of the soft X-ray emission from GC LMXBs \citep{Irwin99,Maccarone03,Maccarone04}. To investigate this, we perform a Spearman Rank test between the four X-ray hardness ratios (listed in table \ref{tab:m31gc_2xmmi}) and the metallicity of a cluster. No strong correlations are observed. The only marginally significant correlation is between HR$_{2}$ and the metallicity, which has a probability of correlation of 92$\%$. To investigate any potential relationships further, we split these data into metal rich ($[Fe/H]>-0.7$) and metal poor clusters. The choice of this split is based on the bimodal peaks in the metallicity identified by \citet{Perrett02}. The fluxes of the X-ray sources, as functions of hardness ratio for the rich and poor clusters, are shown in figure \ref{fig:lmxb_HR_Fx}. The most significant trend is identified between the metal rich and poor clusters was in the HR$_{2}$ (0.5-1keV and 1-2keV). This relationship has a confidence of 99$\%$, but is based primarily on only five metal rich clusters. While significant, the observed relationship is weaker than the relatively strong relationship observed in \textit{ROSAT} observations of these clusters by \citet{Irwin99}. It is possible that this previous relationship may be enhanced due to the difficulty in modelling the background soft excess emission from to M31 in these \textit{ROSAT} data. This could potentially have a greater effect on the more centrally concentrated metal rich clusters. 

\subsection{Stellar collision rate and cluster mass}
\label{sec:xray:lmxb_gamma_mass}

\begin{figure}
 \includegraphics[height=84mm,angle=270]{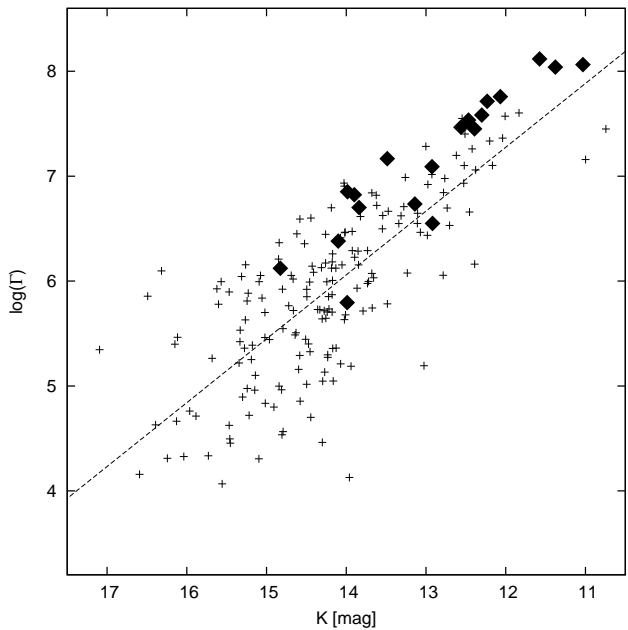}
 \caption{Stellar collision rate vs magnitude for all GCs (crosses) and LMXB hosting clusters (solid diamonds). This shows a clear relationship between the two parameters. This is in good agreement with the predicted relationship $\Gamma\propto M_{tot}^{1.5}$ (dashed line). The LMXB hosting clusters are found to have higher than average collision rates for their magnitude.}
 \label{fig:lmxb}
\end{figure}

In previous work, \citet{Peacock09} demonstrated a relationship between the mass and the stellar collision rate of M31's clusters and the presence of LMXBs. Since this work, significant contamination from non-cluster sources has been removed from the M31 globular cluster catalogues. In this study, we have also presented more homogeneous and complete X-ray data than was previously available. For these reasons we reconsider this relationship, based on these new data. 

Stellar collision rates are available for the 213 old clusters covered by the K-band survey of \citet{Peacock09}. Figure \ref{fig:lmxb_gamma_K} shows the stellar collision rate and K-band luminosity of these clusters. It can be seen that the LMXBs are found to favour both brighter (and hence more massive) clusters and those with higher stellar collision rates. A K-S test demonstrates that these relationships are highly significant with probabilities that they are drawn from the same population of $5\times10^{-4}$ and $4\times10^{-7}$ for luminosity and stellar collision rate, respectively. While still highly significant, the luminosity relationship is slightly weaker than that found by \citet{Peacock09}. This is likely due to the identification of several new X-ray sources in fainter than average clusters. In this study, we have also removed young clusters from our data. This will also weaken the luminosity relationship because all of these young clusters are relatively faint and not associated with any X-ray sources. 

Figure \ref{fig:lmxb} shows the stellar collision rate as a function of K-band luminosity for all GCs studied (crosses) and those containing an LMXB (diamonds). This figure demonstrates that there is a clear relationship between the luminosity of a GC and its stellar collision rate. In order to explore the relative effects of these parameters on LMXBs, we need to consider this relationship. We assume a power law relationship between the mass of a cluster and its stellar collision rate and a K-band mass ($M$) to light ratio of 1, for all clusters. From an unweighted fit to all clusters, in log~space (shown in figure \ref{fig:lmxb}), we find that $\Gamma\propto M_{tot}^{1.53}$. We note that this relationship is consistent with that found for the Milky Way's GCs and with the theoretical approximation that $\Gamma\propto M_{tot}^{1.5}$ \citep{Davies04}. 

\begin{figure}
 \includegraphics[height=84mm,angle=270]{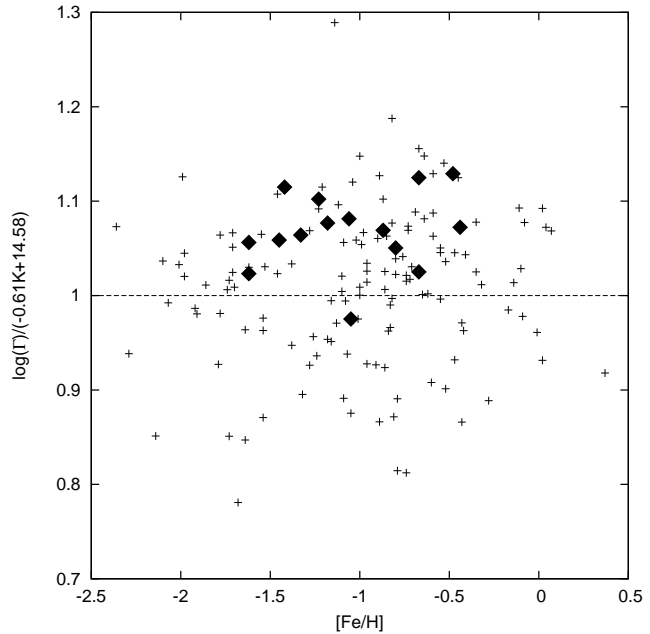}
 \caption{Residual from the $\Gamma$-magnitude relationship (identified in figure \ref{fig:lmxb}) as a function of metallicity. This figure includes only those clusters with metallicity estimates. Symbols are as in figure \ref{fig:lmxb}. It can be seen that the LMXB hosting clusters are offset to higher than average $\Gamma$ for their mass. As demonstrated by this figure, this offset is unlikely to be due to any metallicity effects. }
 \label{fig:gamma_detrend_Z}
\end{figure}

It can be seen from figure \ref{fig:lmxb} that the LMXB hosting clusters have higher than average collision rates for their mass. We investigate whether this is a statistically significant effect by first detrending the data using the derived relationship between $\Gamma$ and mass. We then run a K-S test between all clusters and the LMXB hosting clusters and find a probability of 1$\times$10$^{-4}$ that they are drawn from the same distribution. It should be noted that there is some uncertainty in the actual relationship between $\Gamma$ and mass due to the scatter in figure \ref{fig:lmxb}. To ensure that our results are robust to errors in this relationship, we rerun the tests assuming $\Gamma\propto M_{tot}^{1.25,1.75}$. This results in probabilities of 2$\times$10$^{-6}$ and 1.8$\times$10$^{-3}$ that these clusters are drawn from the same distributions. This demonstrates that, even for the highest reasonable relationship between the mass and collision rate of a cluster, the LMXBs are found to favour those clusters with higher than average stellar collision rates for their mass. This result implies that the stellar collision rate is the primary parameter related to the presence of an LMXB. 

Given that LMXBs are also found to favour metal rich clusters, it is interesting to consider whether cluster metallicity has a significant effect on the relationships we infer from figure \ref{fig:lmxb}. In figure \ref{fig:gamma_detrend_Z} we show the residuals from the derived $\Gamma$-magnitude relationship as a function of metallicity. As discussed above, the LMXB hosting clusters can be seen to have higher than average collision rates for their luminousity. It can be seen that metallicity effects are unable to explain this offset. This is likely due to the fact that the individual effects of both collision rate and luminousity are significantly stronger than the observed metallicity effect. 

\begin{figure}
 \vspace{5mm}
 \includegraphics[height=84mm,angle=270]{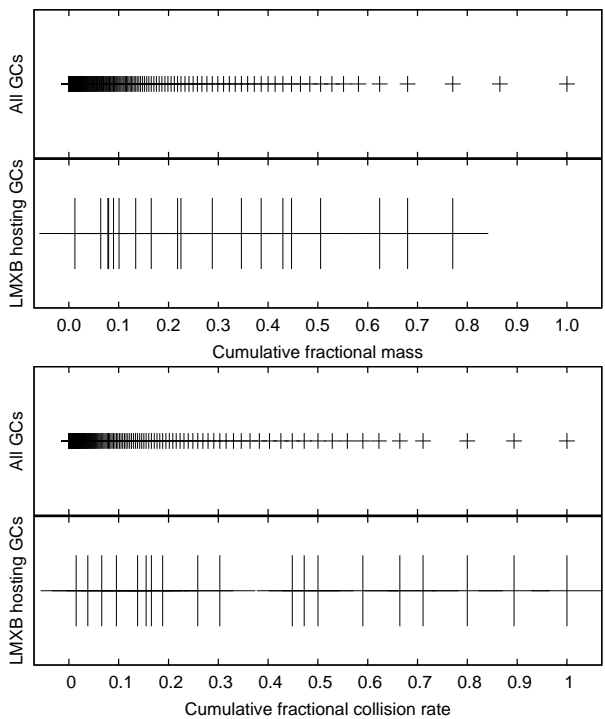}
 \caption{The cumulative fractional mass (top) and cumulative fractional collision rate (bottom) of all GCs in our sample. The top panels show all GCs (small crosses) and the bottom panels show only the GCs known to host LMXBs (large crosses). If the formation of LMXBs is linearly proportional to either stellar collision rate or stellar mass, then the LMXB hosting clusters should be evenly distributed along this plot.}
 \label{fig:reduced-gamma}
\end{figure}

If these LMXBs are formed through dynamical interactions, then the presence of an LMXB should be directly proportional to the stellar collision rate. In order to investigate whether the M31 cluster system is consistent with this, we follow the method of \citet{Verbunt87}. First, we find the fractional stellar collision rate of every cluster by dividing its stellar collision rate by the total collision rate of all of the clusters considered. We then sort the clusters by this fractional collision rate and find the cumulative value for each cluster, such that the cluster values run from 0 to 1. The result of this is that the number of collisions occurring in the GC system should now be evenly distributed between 0 and 1 (i.e. we expect 10$\%$ of the total collisions to occur in the clusters with values in the range 0-0.1). In this way, if the formation of LMXBs is linearly proportional to the stellar collision rate, we expect that their host clusters should be evenly spaced in this plot. Figure \ref{fig:reduced-gamma} shows this plot for both the stellar collision rates and masses of these clusters. It can be seen from this figure that the LMXB hosting clusters are more evenly spread than the total cluster population in both mass and density. This figure also demonstrates that the LMXB hosting clusters are more evenly distributed in stellar collision rate than in mass. 

Figure \ref{fig:reduced-gamma} suggests that there may be a more LMXBs in low collision rate clusters than would be expected based solely on dynamical interactions. A similar overabundance can be seen in the Milky Way's low collision rate clusters \citep[see e.g. figure 1 of][]{Verbunt87}. This may point to additional parameters influencing the formation or evolution of LMXBs in these clusters. However, running a K-S test between these data and a linear relationship, suggests that this bunching is not very significant (with a null hypothesis probability of 12$\%$). We also perform a Wilcoxen-Mann-Whitney test between these data and an evenly distributed dataset, finding a probability of 15$\%$ that the samples are drawn from the same distribution. It should also be noted that, a linear stellar collision rate may predict more than one LMXB in the highest collision rate clusters. Such an effect may result in a non-linear distribution towards the right of figure \ref{fig:reduced-gamma}. We therefore believe that these data are consistent with the formation of LMXBs being linearly proportional to the stellar collision rate. 

\section{Conclusions}
\label{sec:xray:conclusions}

We associate LMXBs with 41 of M31's confirmed old clusters using the 2XMMi catalogue. In addition to these clusters, we identify four other clusters in the literature which have previously been proposed to host LMXBs. Three of the clusters we identify are newly identified and for two other clusters, which were previously identified only in \textit{ROSAT} observations, we confirm their association with the proposed clusters. By using updated optical catalogues of M31's clusters, we show that three of the previously proposed LMXB hosting clusters are now known to be background galaxies. 

LMXBs are identified in $\sim11\%$ of the clusters surveyed. It is likely that the true fraction of LMXB hosting clusters is lower than this because of incompleteness in the current M31 GC catalogues. We confirm the previously identified result that M31's GCs contain more very bright LMXBs than the Milky Way's GCs. This result is likely due to the small number of GC LMXBs in the Milky Way, rather than a difference in the LMXB population. 

By considering the properties of GCs which host LMXBs, the previously proposed relationship between the metallicity of M31's clusters and the presence of an LMXB is identified. This relationship is weaker than, but consistent with, that found in early type galaxies. Our data suggest some evidence for the proposed relationship between metallicity and the hardness ratio of the observed X-ray emission, but this is relatively weak compared with the previously proposed trend. We show highly significant relationships between the presence of an LMXB and both the stellar collision rate and mass of its host GC. The stellar collision rate is found to be the best discriminator in selecting LMXB hosting GCs. We suggest that the weaker relationship between mass and LMXB presence may be primarily due to the relationship between mass and stellar collision rate. Our results demonstrate that the stellar collision rate is likely to be a fundamental parameter related to the formation of LMXBs. This result is in agreement with previous studies of GCs in the Milky Way, M31 and Cen~A. The linear relationship found between the presence of an LMXB and the stellar collision rate is in good agreement with the systems being formed by dynamical interactions. This also suggests that the current dynamical properties of the GCs are related to their current LMXB populations. 

\section*{Acknowledgements}

We would like to thank Pepi Fabbiano, Sean Farrell, Marat Gilfanov, Christian Knigge, Natalie Webb and Zhongli Zhang for helpful discussions. 

\bibliographystyle{mn2e}
\bibliography{bibliography_etal}

\label{lastpage}

\end{document}